\documentclass[aps,prd,amsfonts,amsmath,tightenlines,nofootinbib,aps,11pt]{revtex4}
\usepackage{epsfig,color,eso-pic}
\usepackage{amsmath,amscd,amssymb}
\usepackage{graphics,bigints}
\usepackage{diagbox}

\def\sech{\mathop{\rm sech}\nolimits}
\newcommand{\imu}{{\rm i}}
\newcommand{\fract}[2]{\mbox{\small $\frac{#1}{#2}$}}

\begin{document}

\title{One-Loop Quantum Stress-Energy Tensor for the Kink and
sine-Gordon Solitons}

\author{N. Graham$^{a)}$, H. Weigel$^{b)}$}

\affiliation{
$^{a)}$Department of Physics, Middlebury College
Middlebury, VT 05753, USA\\
$^{b)}$Institute for Theoretical Physics, Physics Department,
Stellenbosch University, Matieland 7602, South Africa}

\begin{abstract}
We compute the renormalized one-loop quantum corrections to
the energy density $T_{00}(x)$ and pressure $T_{11}(x)$ for solitons
in the $1+1$ dimensional scalar sine-Gordon and kink models. We show
how precise implementation of counterterms in dimensional
regularization resolves previously identified discrepancies between
the integral of $T_{00}(x)$ and the known correction to the total energy.
\end{abstract}

\maketitle

\section{Introduction}

In hadron physics, the form factors of the energy-momentum tensor have
recently  attracted broader attention because they can be related to
generalized parton distributions \cite{Ji:1996nm} that are
experimentally accessible \cite{Burkert:2018bqq}.
The empirical status of these form factors has been summarized recently
in a mini-review \cite{Polyakov:2018zvc}. Theoretical input is
available from lattice simulations \cite{Hagler:2009ni} as well as
from calculations in the bag \cite{Ji:1997gm}, Skyrme,
\cite{Cebulla:2007ei} and quark soliton models \cite{Petrov:1998kf}. 
Even though soliton models for baryons in three space dimensions are not 
fully renormalizable, estimates \cite{Meier:1996ng} with  renormalization 
prescriptions guided by chiral perturbation theory \cite{Gasser:1983yg} 
indicate that there are substantial quantum contributions to the baryon mass.  
Similar results should thus also hold for the densities contained in the 
energy-momentum tensor. It is therefore important to gain a full understanding 
of the quantum contributions to the energy-momentum tensor in soliton models. 
As a first step we therefore explore this problem in low-dimensional,
renormalizable models.

Topological solitons in 1+1-dimensional scalar field theory, such as
the $\phi^4$ kink and the sine-Gordon soliton, provide an ideal
theoretical laboratory in which to study the effects of quantum
fluctuations~\cite{Coleman}. Because the resulting potentials for
small-amplitude quantum fluctuations are of the exactly solvable
P\"oschl-Teller form, one can carry out analytic calculations to
obtain exact results at one loop, such as the quantum  correction to
the soliton's energy \cite{Dashen:1974ci}. These calculations often
highlight subtleties of renormalization.  For example, in the
extension to a supersymmetric model, one might expect the quantum
correction to the soliton's energy to be zero, since the spectra of
Bose and Fermi modes are identical except for zero modes, and enter
with opposite signs.  However, by Levinson's theorem, the mismatch
between zero modes also implies a difference in the phase shifts for
the quantum fluctuations about the soliton.  Since these phase shifts
parameterize the change of the density of states generated by the
soliton, there is a nonzero correction to the energy.  Because this
correction is negative, it appears to violate the BPS bound, but the
central charge receives an equal correction by the same mechanism
\cite{super1d}.  This result reflects the presence of an anomaly in
the supersymmetric theory \cite{SVV}, an example of the more general
relationship between anomalies and Levinson's Theorem \cite{Niemi}.

Recent work \cite{PhysRevD.107.065002,Ito:2023oby} studying local densities 
has highlighted another such subtlety of these quantum corrections, for the 
case of local densities in a scalar theory. Here again the subtleties arise 
from the zero mode. Since it corresponds to translation of the soliton, it 
must be quantized as a collective coordinate rather than as a small-amplitude
quantum fluctuation. But when one carries out this
calculation for the one-loop quantum correction to the energy density,
it appears that the integral over $x$ of the result does not agree
with the known correction to the total energy. To address this
discrepancy, Ref.~\cite{PhysRevD.107.065002} introduces an additional
counterterm for the stress-energy tensor, while Ref.~\cite{Ito:2023oby} 
obtains a spatially constant additional correction. However, because the 
former is an \emph{ad hoc} modification of the original theory and the 
latter represents a redefinition of the renormalization condition for the 
cosmological constant, both of these modifications are expected to change 
the total energy as well. In the following, we argue that this puzzle is
instead resolved by a subtle aspect of the renormalization process.  When
the counterterm is specified precisely in dimensional regularization,
one obtains a result that agrees with the known total energy with
no need for additional modifications.

We begin by reviewing the scalar field theory models and their soliton
solutions in Sec.~\ref{sec:model}.  Then in Sec.~\ref{sec:spectral} we
introduce the quantum corrections, obtained using spectral methods in
dimensional regularization, leading to analytic results for both
models in Sec.~\ref{sec:results}, and we include additional supporting
calculations in two Appendices.

\section{Model}
\label{sec:model}

We consider a scalar model defined by the Lagrangian density 
\begin{equation}
{\cal L} = \frac{1}{2} \dot \phi^2 - \frac{1}{2} {\phi'}^2 - U(\phi)
\end{equation}
in $D=1+1$ spacetime dimensions, with dot and prime denoting time and
space derivatives of $\phi$, respectively. The equation of motion is
\begin{equation}
\ddot{\phi}^2 = \phi'' - U'(\phi) \,,
\end{equation}
where $U'(\phi)$ refers to the derivative of $U$ with respect to its
argument.  Then a static solution $\phi_0(x)$ obeys
\begin{equation}
\frac{d\phi_0(x)}{dx} = \pm \sqrt{2 U(\phi_0(x))}
\label{eq:soliton}
\end{equation}
for the soliton and antisoliton.  We will consider both the
sine-Gordon soliton, for which
\begin{equation}
U^{\rm (SG)}(\phi) = m^2 (1-\cos \phi)
\quad \hbox{and} \quad
\phi_0^{\rm (SG)}(x) = 4 \arctan e^{\pm mx},
\end{equation}
and the $\phi^4$ kink, for which 
\begin{equation}
U^{\rm (kink)}(\phi) = \frac{m^2}{8} \left(\phi^2 - 1\right)^2
\quad \hbox{and} \quad
\phi_0^{\rm (kink)}(x) = \pm \tanh\frac{mx}{2}\,,
\label{eq:Ukink}
\end{equation}
where in both cases $m$ is the mass of perturbative fluctuations. 

In a fuller analysis of the theory, the Lagrangian would include 
an overall factor of $\fract{m^2}{2\lambda}$, which does not affect the
equations of motion.  It tracks the order of the loop expansion in
terms of the four-point coupling constant $\lambda$, with the classical
contribution entering at order $\lambda^{-1}$ and the one-loop
contribution that we focus on here entering at order $\lambda^{0}$.  In
that analysis, it is convenient to introduce the unscaled boson
field $\varphi=\frac{m}{\sqrt{2\lambda}}\phi$.  We will use this
scaling in Appendix~\ref{sec:eta}, where we carry out a perturbative
expansion in the coupling.

Expanding to quadratic order, we obtain the equation for small oscillations 
$\eta(x,t)=\eta_k(x){\rm e}^{-\imu\omega t}$ around the soliton,
\begin{equation}
-\frac{d^2 \eta_k(x)}{dx^2} + V(x) \eta_k(x) = k^2 \eta_k(x)\,,
\label{eq:smallosc}
\end{equation}
where $\omega = \sqrt{k^2+m^2}$ is the mode frequency
and the small-oscillation potential is given by
\begin{equation}
V(x) = U''(\phi_0(x)) - m^2 = -\frac{\ell+1}{\ell} m^2 \sech^2 \frac{mx}{\ell}
\end{equation}
with $\ell=1$ for the sine-Gordon soliton and $\ell=2$ for the kink.

Both potentials are reflectionless and are exactly solvable. The
continuum modes are
\begin{equation}
\eta^{\rm (SG)}_k(x) = \frac{\imu}{\sqrt{k^2+m^2}}
\left(k+\imu m \tanh mx \right) {\rm e}^{\imu kx}
\end{equation}
for the sine-Gordon soliton and
\begin{equation}
\eta^{\rm (kink)}_k(x) = \frac{1}{\sqrt{k^2+m^2}\sqrt{k^2+\frac{m^2}{4}}}
\left(\frac{m^2}{4} + k^2 + \frac{3}{2} \imu m k \tanh \frac{mx}{2}
- \frac{3}{4} m^2 \tanh^2 \frac{mx}{2}\right) {\rm e}^{\imu kx}
\end{equation}
for the kink. These continuum modes are normalized such that
$\lim_{x\to\pm\infty}|\eta_k(x)|=1$. Both models have zero modes with
$\omega = 0$,
\begin{equation}
\eta^{\rm (SG)}_0(x) = \sqrt{\frac{m}{2}} \sech mx
= \frac{1}{\sqrt{8m}} \frac{d}{dx} \phi^{\rm (SG)}_0(x)
\end{equation}
and
\begin{equation}
\eta^{\rm (kink)}_0(x) = \sqrt{\frac{3m}{8}} \sech^2 \frac{mx}{2}
= \sqrt{\frac{3}{2m}} \frac{d}{dx} \phi^{\rm (kink)}_0(x) \,.
\end{equation}
The kink also has a ``shape'' mode
\begin{equation}
\eta^{\rm (kink)}_1(x) = \sqrt{\frac{3m}{4}}
\tanh \frac{mx}{2} \sech \frac{mx}{2}
\end{equation}
with frequency $\displaystyle \omega = \sqrt{\frac{3}{4}}m$.
We use standard normalization for the bound state wave-functions,
$\int_{-\infty}^\infty dx |\eta_i(x)|^2=1$.

We define the Green's function with outgoing wave boundary conditions,
\begin{equation}
G(x_1,x_2,k) = \frac{\imu}{2k} \eta_{-k}(x_<) \eta_{k}(x_>) \,,
\end{equation}
where $x_<$ $(x_>)$ is the smaller (larger) of $x_1$ and $x_2$, along
with the corresponding free Green's function
\begin{equation}
G_0(x_1,x_2,k) = \frac{\imu }{2k} e^{-\imu kx_<} e^{\imu kx_>} = 
\frac{\imu}{2k} e^{\imu k|x_1-x_2|}\,.
\end{equation}

\section{Spectral Method}
\label{sec:spectral}

We wish to compute one-loop quantum corrections to the stress-energy
tensor. By symmetry, its off-diagonal components vanish, so we only need
to calculate $T_{00}$ and $T_{11}$.

Following the approach of Ref.~\cite{Ito:2023oby}, we express the
calculation of the stress-energy tensor
\begin{equation}
T_{\mu \nu} = \partial_\mu \phi \partial_\nu \phi - g_{\mu \nu} {\cal L}
\label{eq:Tmunu}
\end{equation}
in terms of four components $T_1(x)$, $T_2(x)$, $T_3(x)$, and $T_4(x)$. In 
this decomposition, $T_1(x)$ and $T_2(x)$ are the contributions from the 
quadratic field fluctuations, while $T_3(x)$ and $T_4(x)$ are the 
contributions from linear fluctuations that are of the same order in 
$\lambda$ as $T_1(x)$ and $T_2(x)$, {\it cf.~} Appendix \ref{sec:eta}. 
Furthermore $T_1(x)$ and $T_3(x)$ are the derivative terms,
which enter with the same sign in both $T_{00}$ and $T_{11}$,
while $T_2(x)$ and $T_4(x)$ are the potential terms, which enter with
opposite signs.

We begin by considering the quadratic contribution to $T_{00}$,
\begin{eqnarray}
T_1(x) + T_2(x) &=& \frac{1}{2}\langle \dot \eta(x,t)^2 \rangle +
\frac{1}{2}\langle {\eta^\prime(x,t)}^2 \rangle + V(x)\langle
\eta(x,t)^2 \rangle\cr
&=& \langle \dot \eta(x,t)^2 \rangle + \frac{1}{4}
\left\langle \frac{d^2}{dx^2}\left(\eta(x,t)^2\right) \right\rangle \,,
\label{eqn:energydensity}
\end{eqnarray}
where angle brackets refer to renormalized expectation values with the
zero mode contribution omitted, and we have used the equations of
motion to write the same quantity in two different forms.

We will first apply the spectral method \cite{density,Graham:2002yr}
to the second line of Eq.~(\ref{eqn:energydensity}).  In order to
implement the renormalization counterterms precisely, we use
dimensional regularization and introduce $n$ transverse dimensions
\cite{interfac}, so that the soliton becomes a domain wall in $n+1$
space dimensions.  We obtain
\begin{align}
T_1(x) + T_2(x) &=
 -\frac{\Gamma\left(-\tfrac{n+1}{2}\right)}{(4\pi)^{\frac{n+1}{2}}} 
\Bigg\{ \sum_{\hbox{\tiny bound states $j$}} 
\frac{1}{2}\left(\omega_j^{n+1} - m^{n+1}\right) 
\left(\eta_j(x)^2 + \frac{n+1}{4\omega_j^2}
\frac{d^2}{dx^2}\left(\eta_j(x)^2\right) \right) \cr
&\hspace{1cm}+ \int_0^\infty \frac{dk}{2\pi}
\left((k^2+m^2)^{\frac{n+1}{2}} - m^{n+1}\right) \frac{2k}{\imu} \cr
&\hspace{2cm} \times \Bigg[G(x,x,k) - G_0(x,x,k) \left(1+
\frac{V(x)}{2k^2}\right) \cr
&\hspace{3cm} + \frac{n+1}{4(k^2+m^2)} \frac{d^2}{dx^2}G(x,x,k) 
+ \frac{\imu m(n+1)}{4k(k^2+m^2)^2}
\frac{d^2}{dx^2} \left(\eta_0(x)^2\right)
\Bigg] \Bigg\}\cr
&= -\frac{1}{2 (4\pi)^{\frac{n+1}{2}}
\Gamma\left(\tfrac{n+3}{2}\right)} 
\int_m^\infty d\kappa \left(\kappa^2 - m^2\right)^{\frac{n+1}{2}}
2\kappa \cr &\hspace{1cm} 
\times \Bigg[G(x,x,\imu\kappa) - G_0(x,x,\imu\kappa) \left(1-
\frac{V(x)}{2\kappa^2}\right) \cr
&\hspace{2cm} + \frac{n+1}{4(m^2-\kappa^2)} \frac{d^2}{dx^2}G(x,x,\imu\kappa) 
+ \frac{m(n+1)}{4\kappa(m^2-\kappa^2)^2}
\frac{d^2}{dx^2} \left(\eta_0(x)^2\right)
\Bigg] \,.
\label{eqn:NPB}
\end{align}
There are two terms involving the free Green's function. The first one,
just $G_0(x,x,k)$, subtracts the vacuum contribution and defines the 
zero of energy. The second one, proportional to $G_0(x,x,k)V(x)$,
is the counterterm contribution, {\it cf.} Eq.~(\ref{eq:ct1}) in 
Appendix~\ref{sec:eta},
\begin{equation}
\Delta T_{\mu\nu}=-g_{\mu\nu}\Delta \mathcal{L}\,.
\label{Tmunucounter}
\end{equation}
These counterterms are defined precisely and unambiguously via dimensional 
regularization \cite{Farhi:2000ws}. For the models under consideration, they
are \cite{deVega:1976sm}
\begin{equation}
\Delta \mathcal{L}^{\rm (SG)}=C^{\rm (SG)}\left(1-\cos\phi\right)
\qquad {\rm and}\qquad
\Delta \mathcal{L}^{\rm (kink)}=C^{\rm (kink)}\left(\phi^2-1\right)\,.
\label{Lcounter}
\end{equation}
The coefficients are chosen such that the tadpole diagram is exactly
canceled. As a result, we obtain a fully continuum formulation without
a need for any discretization or additional counterterms. For real
$k$, we have used the completeness relation
\begin{equation}
\sum_{\hbox{\tiny bound states $j$}} 
\eta_j(x)^2 + \int_0^\infty \frac{dk}{\pi} \frac{2k}{\imu}
\left[G(x,x,k) - G_0(x,x,k)\right] = 0\,,
\end{equation}
which is the local equivalent of Levinson's theorem, to replace $\omega$ 
by $\omega -m$ in both the continuum and bound state contributions, as is 
necessary to obtain convergence of the integral at small $k$. We stress that 
the above sums over bound states include the zero mode. The last term in 
Eq.~(\ref{eqn:NPB}) removes the contribution from the zero mode, since 
it should be excluded in this calculation. Then we have used contour
integration to write the integral on the imaginary axis
$k=i\kappa$. The poles enclosed by this contour exactly cancel the
explicit contributions from the bound states.

Next we use the equations of motion to recast this result into the
form of the first line of Eq.~(\ref{eqn:energydensity}). We obtain
\begin{align}
T_1(x) + T_2(x) &= -\frac{1}{2 (4\pi)^{\frac{n+1}{2}}
\Gamma\left(\tfrac{n+3}{2}\right)} 
\int_m^\infty d\kappa \left(\kappa^2 - m^2\right)^{\frac{n+1}{2}}
2\kappa\frac{(n+1)}{2(m^2-\kappa^2)} 
\label{eqn:quadratic}\\
&\hspace{1cm} \times \Bigg[\frac{1-n}{1+n}(m^2-\kappa^2)
\left(G(x,x,\imu \kappa) - G_0(x,x,i\kappa)\right)\cr
&\hspace{1.5cm} +\frac{d}{dx}\frac{d}{dy}
\left(G(x,y,\imu \kappa) - G_0(x,y,\imu \kappa)\right)\big|_{y=x}
+ \frac{m}{\kappa (m^2-\kappa^2)} \left(\frac{d\eta_0(x)}{dx}\right)^2  \cr
&\hspace{1.5cm} + V(x) \left(G(x,x,\imu \kappa) - G_0(x,x,\imu \kappa)
\frac{\kappa^2-m^2}{\kappa^2(n+1)}\right) \cr
&\hspace{1.5cm} + m^2 \left(G(x,x,\imu \kappa) - G_0(x,x,\imu \kappa)\right)
+ \frac{m}{\kappa (m^2-\kappa^2)} (V(x)+m^2)\eta_0(x)^2
\Bigg], \nonumber
\end{align}
which we emphasize is simply an algebraic reorganization of
Eq.~(\ref{eqn:NPB}) using the small amplitude fluctuation equation
(\ref{eq:smallosc}).

The key expression to focus on here is the factor of 
$\frac{\kappa^2-m^2}{\kappa^2(n+1)}$ multiplying the free Green's
function in the potential term.  The need for this factor would not be
apparent from the original expression, since one would just expect
to subtract the free Green's function; this factor approaches unity at
large $\kappa$ and $n=0$, but its difference from unity makes a finite
contribution to the energy density, yielding an expression with the
correct integral over space, the well-established vacuum 
polarization energy (VPE), {\it cf.} Eq.~(\ref{eq:VPE}) below.  
Because the spectral method provides the
exact counterterm in dimensional regularization, this factor is
unambiguously required in order for the resulting expression to be
consistent with Eq.~(\ref{eqn:NPB}).

To compute $T_{11}$, we will also need the difference $T_1(x) -
T_2(x)$ between the derivative and potential terms.  The only subtlety
here is how to divide the counterterm contribution described above
between the two terms.  We make this determination
by requiring that the contribution linear in $V(x)$ vanish.  With
this condition, we obtain
\begin{align}
T_1(x) - T_2(x) &= -\frac{1}{2 (4\pi)^{\frac{n+1}{2}}
\Gamma\left(\tfrac{n+3}{2}\right)} 
\int_m^\infty d\kappa \left(\kappa^2 - m^2\right)^{\frac{n+1}{2}}
2\kappa\frac{(n+1)}{2(m^2-\kappa^2)} 
\label{eqn:quadratic2}\\
&\hspace{1cm} \times \Bigg[\frac{1-n}{1+n}(m^2-\kappa^2) 
\left(G(x,x,\imu \kappa) - G_0(x,x,i\kappa)\right)\cr
&\hspace{1.5cm} +\frac{d}{dx}\frac{d}{dy}
\left(G(x,y,\imu \kappa) - G_0(x,y,\imu \kappa)\right)\big|_{y=x}
+ \frac{m}{\kappa (m^2-\kappa^2)} \left(\frac{d\eta_0(x)}{dx}\right)^2  \cr
&\hspace{1.5cm} - V(x) \left(G(x,x,\imu \kappa) - G_0(x,x,\imu \kappa)
\frac{\kappa^2-2m^2}{\kappa^2(n+1)}\right) \cr
&\hspace{1.5cm} - m^2 \left(G(x,x,\imu \kappa) - G_0(x,x,\imu \kappa)\right)
- \frac{m}{\kappa (m^2-\kappa^2)} (V(x)+m^2)\eta_0(x)^2
\Bigg] \,. \nonumber
\end{align}
In this expression, we note the factor of 
$\frac{\kappa^2-2 m^2}{\kappa^2(n+1)}$ in the counterterm contribution needed 
to implement the renormalization condition, where one might have expected 
the same factor as in Eq.~(\ref{eqn:quadratic}), without the factor of $2$. 
This difference arises as a result of contributions to the integrand that 
formally vanish at $n=0$, but which give a finite, nonzero contribution to the
integral in dimensional regularization, and can be viewed as a
consequence of the trace anomaly \cite{PhysRevD.15.1469}.  With this
choice, we obtain the  correct contribution to the trace
$T_{\mu}^{\hspace{2mm}\mu}$ due to the soliton,
\begin{equation}
2 T_2(x) = \frac{1}{4\pi} V(x) + \langle \eta(x)^2 \rangle 
\left(V(x)+m^2\right)\,,
\end{equation}
where the first term on the right-hand side reflects the contribution
from the anomaly, and we have computed the renormalized expectation value 
$\langle \eta(x)^2 \rangle$ in Appendix~\ref{sec:etasq}. The soliton's  
translational variance yields $\partial_\mu T^{\mu 1}\ne0$, so the
composite operator $T^{\mu 1}$ is not protected against further
renormalization.

\section{Results}
\label{sec:results}

We can now carry out the integrals using {\tt Mathematica} and take
the limit $n\to 0$.  We obtain
\begin{eqnarray}
T_1^{\rm (SG)}(x) + T_2^{\rm (SG)}(x) &=& 
-\frac{m^2 \sech^2 mx}{2 \pi }\,,
\cr
T_1^{\rm (SG)}(x) - T_2^{\rm (SG)}(x) &=& 0
\end{eqnarray}
for the sine-Gordon model and
\begin{eqnarray}
T_1^{\rm (kink)}(x) + T_2^{\rm (kink)}(x) &=& 
m^2 \left(\frac{1}{4 \sqrt{3}}-\frac{3}{8 \pi }\right)
\sech^2\frac{m x}{2}-\frac{17 m^2}{32 \sqrt{3}} \sech^4\frac{m x}{2}
+\frac{5 m^2}{16 \sqrt{3}}  \sech^6\frac{m x}{2}\,,
\cr
T_1^{\rm (kink)}(x) - T_2^{\rm (kink)}(x) &=&
\frac{\sqrt{3} m^2 }{32} \sech^4 \frac{mx}{2}
-\frac{m^2}{16 \sqrt{3}} \sech^6\frac{mx}{2}
\end{eqnarray}
for the kink.

In contrast to the total energy, the densities have contributions linear in 
the quantum fluctuations
\begin{equation}
T^{\rm (linear)}_{\mu\mu}=2\partial_\mu\eta\partial_\mu\phi_0
-g_{\mu\mu}\left[\partial_\nu\eta\partial^\nu\phi_0-U^\prime(\phi_0)
\eta\right]\,,
\label{eq:Tlinear}\end{equation}
where the index $\mu$ is not summed. Following
Ref.~\cite{Ito:2023oby} and as outlined
in Appendix~\ref{sec:eta}, we compute the expectation value
\begin{align}
\langle \eta(x) \rangle &= 
-\frac{1}{2}\int_{-\infty}^\infty dy \, U'''(\phi_0(y))
\int_0^\infty dq \, \frac{1}{2 \pi \sqrt{q^2+m^2}} 
\frac{2q}{\imu}\left(G(y,y,q)-G_0(y,y,q)\right) \cr
&\hspace{2cm}\times \int_0^\infty dk \,
\frac{1}{\pi(k^2+m^2)} \frac{2k}{\imu}G(x,y,k)\,,
\label{eq:eta}\end{align}
which in turn gives the contributions from $T_{\mu\mu}^{\rm (linear)}$
to the densities
\begin{equation}
T_3(x) = \phi_0'(x) \frac{d}{dx}\langle \eta(x) \rangle 
\qquad {\rm and}\qquad
T_4(x) = U'(\phi_0(x)) \langle \eta(x) \rangle \,.
\end{equation}
Explicit calculations yield
\begin{align}
T_3^{\rm (SG)}(x) &= -\frac{m^2}{4 \pi} \sech^2 mx +\frac{m^2}{2 \pi}
\sech^4 mx\,, \cr
T_4^{\rm (SG)}(x) &= -\frac{m^2}{4 \pi} \sech^2 mx +\frac{m^2}{4 \pi}
\sech^4 mx
\end{align}
for the sine-Gordon model and  \cite{Ito:2023oby}
\begin{align}
T_3^{\rm (kink)}(x) &=
m^2\left(\frac{\sqrt{3}}{32} m x \tanh\frac{m x}{2}
-\frac{7}{32 \sqrt{3}}-\frac{3}{16 \pi}\right)\sech^4 \frac{m x}{2} 
+ m^2\left(\frac{\sqrt{3}}{16}+\frac{9}{32 \pi}\right)
\sech^6 \frac{mx}{2}\,,\cr
T_4^{\rm (kink)}(x) &=
m^2 \left(\frac{\sqrt{3}}{32} m x \tanh\frac{mx}{2}
-\frac{1}{8 \sqrt{3}}-\frac{3}{16 \pi}\right) \sech^4 \frac{m x}{2}
+ m^2\left(\frac{1}{8 \sqrt{3}}+\frac{3}{16 \pi }\right)
\sech^6 \frac{m x}{2}
\end{align}
for the kink. By construction of the soliton, Eq.~(\ref{eq:soliton}),
we have $\int_{-\infty}^{\infty} dx\left[T_3(x)+T_4(x)\right]=0$ as a
consistency check.

\begin{figure}
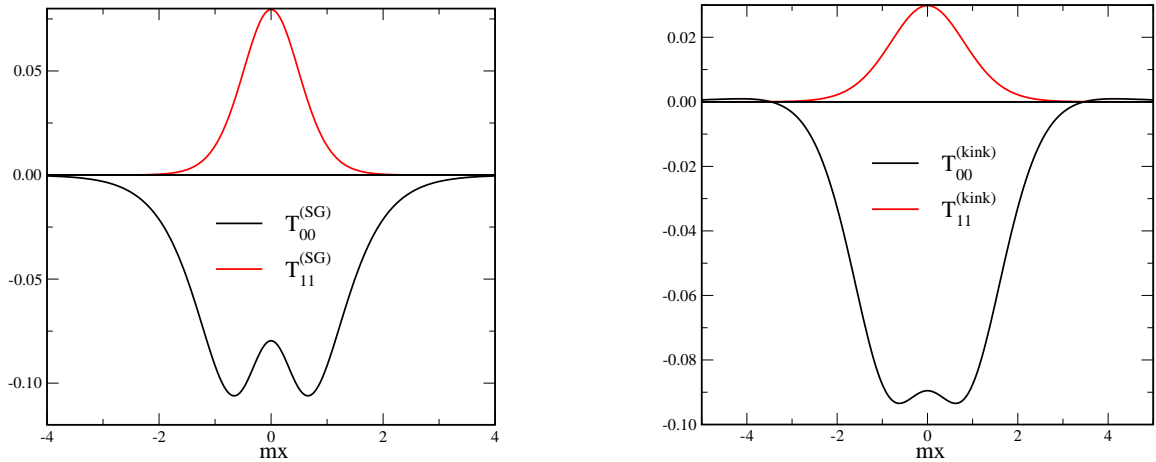

\centerline{
\includegraphics[width=0.4\linewidth]{SG}\hspace{2cm}
\includegraphics[width=0.4\linewidth]{kink}}
\caption{Quantum corrections to the energy and pressure densities
$T_{00}(x)$ and  $T_{11}(x)$, respectively, for the sine-Gordon
(left panel) and kink (right panel) solitons, shown in units of the
perturbative mass squared,~$m^2$.}
\label{fig:densities}
\end{figure}

Putting these results together using
\begin{align}
T_{00}(x) &= T_1(x) + T_2(x) + T_3(x) + T_4(x)\,, \cr
T_{11}(x) &= T_1(x) - T_2(x) + T_3(x) - T_4(x)
\end{align}
we obtain
\begin{align}
T_{00}^{\rm (SG)}(x) &= 
-\frac{m^2}{\pi} \sech^2 mx +\frac{3 m^2}{4 \pi}  \sech^4 mx\,, \cr
T_{11}^{\rm (SG)}(x) &= \frac{m^2}{4 \pi}  \sech^4 mx
\end{align}
for the sine-Gordon model and
\begin{align}
T_{00}^{\rm (kink)}(x) &=  
m^2 \left(\frac{1}{4 \sqrt{3}}-\frac{3}{8 \pi }\right)
\sech^2\frac{m x}{2}+
m^2\left(\frac{5}{8 \sqrt{3}}+\frac{15}{32 \pi }\right)
\sech^6\frac{mx}{2} \cr &\hspace{2cm}
+ m^2 \left(\frac{\sqrt{3}}{16} mx \tanh \frac{m x}{2}
-\frac{7}{8 \sqrt{3}}-\frac{3}{8 \pi}\right) \sech^4\frac{mx}{2}\,, \cr
T_{11}^{\rm (kink)}(x) &=
\frac{3 m^2}{32 \pi} \sech^6\frac{m x}{2}
\end{align}
for the kink. These functions are shown in Fig.~\ref{fig:densities}. 
Like Ref.~\cite{PhysRevD.107.065002} but unlike Ref.~\cite{Ito:2023oby},
we find that $|T_{00}|$ has a local minimum at the center of the soliton. 
However, this minimum is not zero for the sine-Gordon model.

From the above densities we obtain the total energies
\begin{equation}
\int_{-\infty}^\infty dx \, T_{00}^{\rm (SG)}(x) = -\frac{m}{\pi}
\qquad{\rm and}\qquad
\int_{-\infty}^\infty dx \, T_{00}^{\rm (kink)}(x) = 
m \left(\frac{1}{4\sqrt{3}} - \frac{3}{2\pi}\right)
\label{eq:VPE}\end{equation}
in the no-tadpole renormalization scheme, which are the 
well-established results \cite{Dashen:1974ci,Ra82,Graham:2009zz}.
Similarly, the integrated pressures are
\begin{equation}
\int_{-\infty}^\infty dx \, T_{11}^{\rm (SG)}(x) = \frac{m}{3\pi} 
\qquad{\rm and}\qquad
\int_{-\infty}^\infty dx \, T_{11}^{\rm (kink)}(x) =
\frac{m}{5\pi} \,.
\end{equation}

\section{Conclusions}

We have explained calculations of the renormalized corrections to
$T_{00}(x)$ and $T_{11}(x)$ for solitons in the sine-Gordon and
$\phi^4$ kink models, with counterterms precisely specified through
dimensional regularization.  The theory is defined in the continuum,
without the need for artificial boundary conditions, and all
counterterms are fully specified by the no-tadpole renormalization
condition.  When written in terms of quadratic expectation
values, the tadpole graph counterterm enters with a kinematic factor
that would otherwise be unexpected. Incorporating this term gives a
result for $T_{00}(x)$ whose integral gives the correct total energy,
resolving a discrepancy observed in previous studies.  A similar
analysis of $T_{11}(x)$ shows that this calculation
also captures the effects of the trace anomaly. Overall, this result
provides a valuable example of the subtleties of renormalization in
the context of exactly solvable models.

\section{Acknowledgments}
N.\ G.\ is supported in part by the
National Science Foundation (NSF) through grant PHY-2209582.
H.\ W.\ is supported in part by the National Research Foundation of
South Africa (NRF) by grant~150672.

\appendix
\section{Renormalization of $\mathbf{\langle \eta(x)\rangle}$ 
in the kink model}
\label{sec:eta}

In this appendix we describe the renormalization of $\langle \eta(x)\rangle$
in kink model. At the end we will write it in a general form that also
applies to the sine-Gordon model.

Our starting point is the original Lagrangian in terms of the unscaled
field $\varphi=\frac{m}{\sqrt{2\lambda}}\phi$,
\begin{equation}
\mathcal{L}=\frac{1}{2}\partial_\mu\varphi\partial^\mu\varphi-U(\varphi)
\qquad {\rm with}\qquad
U(\varphi)=\frac{\lambda}{4}\left[\varphi^2-\frac{m^2}{2\lambda}\right]^2\,,
\label{eq:lag} \end{equation}
with the $\lambda$ coupling constant explicitly included to allow for
a perturbation expansion. The Lagrangian in Eq.~(\ref{eq:Ukink}) is
obtained by taking $\lambda=\frac{m^2}{2}$.  The counterterm Lagrangian at
one-loop is
\begin{equation}
\mathcal{L}_{\rm ct}=c\lambda \left[\varphi^2-\frac{m^2}{2\lambda}\right]
=\frac{c}{3}\left[U^{\prime\prime}(\varphi)-m^2\right]\,
\stackrel{\varphi\,\to\,\varphi_0}{\mbox{\Large $\longrightarrow$}}
\,\frac{c}{3}V(x)\,,
\label{eq:lagct} \end{equation}
where the coupling constant is factored out to simplify subsequent
expressions.

In the no-tadpole scheme, we have to remove the term linear in $V$
from the effective action
\begin{align}
&\int d^2x \left[\mathcal{L}+\mathcal{L}_{\rm ct}\right]
+\frac{\imu}{2}{\rm Tr}{\rm log}\left[\partial_x^2+m^2+V(x)\right]
=\ldots+\int d^2x\, V(x)\left[\frac{c}{3}
-\frac{\imu}{2}\int \frac{d^2k}{(2\pi)^2}\frac{1}{k^2-m^2}\right]\cr
&\hspace{2cm}
\Longrightarrow\qquad
c=\frac{3\imu}{2}\int \frac{d^2k}{(2\pi)^2}\frac{1}{k^2-m^2}
=\frac{3\imu}{2}\Delta_F^{(0)}(0)\,.
\label{eq:ct1}\end{align}
Here we implicitly take all loop-integrals to be dimensionally regularized.
The free Feynman propagator is defined so that
$\left(\partial^2_x+m^2\right)\Delta_F^{(0)}(x)=-\delta^{2}(x)$. Below
we will also encounter the Feynman propagator of the full theory, which solves
$\left(\partial_x^2+m^2+V(x)\right)\Delta_F(x,y)=-\delta^{2}(x-y)$.

As in the case of the VPE, we expect the densities $T^{\mu\nu}$ to be
$\mathcal{O}\left(\lambda^0\right)$.
Since $\varphi_0=\mathcal{O}\left(1/\sqrt{\lambda}\right)$ we must
therefore collect all terms that contribute to $\langle\eta(x)\rangle$
at $\mathcal{O}\left(\sqrt{\lambda}\right)$.
To this end we need to compute the path integral 
\begin{align}
\langle \eta(x)\rangle&=\frac{1}{Z}\int D\eta\,\eta(x)\,
{\rm e}^{\frac{\imu}{2}\int \eta \Delta_F^{-1}\eta
+\imu\lambda\int d^2z \left[2c\varphi_0\eta-\varphi_0\eta^3\right]}\cr
&=\frac{\imu\lambda}{Z}\int D\eta\,\eta(x)
\int d^2z\, \varphi_0(z) \left[2c\eta(z)-\eta(z)^3\right]
{\rm e}^{\frac{\imu}{2}\int \eta \Delta_F^{-1}\eta}
+\ldots\cr
&=3\imu\lambda\int d^2z\, \varphi_0(z)\Delta_F(x,z)
\left[\Delta_F(z,z)-\Delta^{(0)}_F(0)\right]\,,
\label{eq:expect1} \end{align}
where we have used $\langle \eta(x)\eta(z)\rangle=\imu\Delta_F(x,z)$,
Wick's theorem for 
$\langle \eta(x)\eta(z)^3\rangle$, and Eq.~(\ref{eq:ct1}) for the
counterterm coefficient. 
Note that the cubic term arises from expanding $U(\varphi_0+\eta)$, so
that the coefficient is $U^{\prime\prime\prime}(\varphi_0)/6$. The
linear term originates from the first-order expansion of the
counterterm, which by itself is proportional to
$U^{\prime\prime}(\varphi_0)$, up to an additive constant. Hence the
general form of the expectation value is
\begin{equation}
\langle \eta(x)\rangle=\frac{\imu}{2}\int d^2z\,
U^{\prime\prime\prime}(\varphi_0)
\Delta_F(x,z)\left[\Delta_F(z,z)-\Delta^{(0)}_F(0)\right]\,.
\label{eq:expect2} \end{equation}
For static $\varphi_0$, the integrals over the frequencies in the
propagators and the time coordinate $z_0$ are straightforward,
resulting in Eq.~(\ref{eq:eta}).

\section{Expectation value of $\mathbf{\langle \eta(x)^2\rangle}$}

\label{sec:etasq}

The expectation value of $\langle \eta(x)^2\rangle$ can be
computed by the same methods as we have used above for $T_{\mu\nu}$,
providing a simpler example of the subtleties of renormalization.
The relevant expression is similar to Eq.~(\ref{eqn:NPB}),
but without the derivative terms and with two fewer powers of 
$\omega = \sqrt{m^2 - \kappa^2}$,
\begin{align}
\langle \eta(x)^2\rangle &=
\frac{1}{(4\pi)^{\frac{n+1}{2}}
\Gamma\left(\tfrac{n+3}{2}\right)} 
\int_m^\infty d\kappa \left(\kappa^2 - m^2\right)^{\frac{n-1}{2}} \kappa 
\Bigg[G(x,x,\imu\kappa) - G_0(x,x,\imu\kappa)
+ \frac{m}{\kappa(m^2-\kappa^2)}
\eta_0(x)^2
\Bigg],
\label{eqn:NPB2}
\end{align}
which becomes
\begin{equation}
\langle \eta^{\rm (SG)}(x)^2\rangle = 0
\qquad{\rm and}\qquad
\langle \eta^{\rm (kink)}(x)^2\rangle = 
\frac{1}{4 \sqrt{3}} \sech^2 \frac{m x}{2}\tanh^2 \frac{m x}{2}
\end{equation}
for the two models. Direct calculation shows that
\begin{equation}
-\frac{1}{2} \frac{dV(x)}{dx} \langle \eta(x)^2\rangle
= \frac{d}{dx} \left(T_1(x) - T_2(x)\right)
\label{eq:B3}
\end{equation}
in both models.  Letting  $\eta=\eta(x,t)=\eta_k(x){\rm e}^{-\imu\omega t}$, 
from Eq.~(\ref{eq:Tmunu}) we have
\begin{align}
T_1-T_2&=\frac{1}{2}\left\langle 
\left(\frac{\partial \eta}{\partial t}\right)^2 
+\left(\frac{\partial \eta}{\partial x}\right)^2-U^{\prime\prime}(\phi_0)
\eta^2 \right\rangle\\
\Longrightarrow\qquad 
\frac{d}{dx} \left(T_1-T_2\right)&=
-\left\langle\frac{\partial \eta}{\partial x}\left(
\frac{\partial^2 \eta}{\partial t^2}-\frac{\partial^2 \eta}{\partial x^2}
+U^{\prime\prime}(\phi_0)\eta\right)\right\rangle\\
&\hspace{2cm}
+\left\langle\frac{\partial}{\partial t}\left(\frac{\partial \eta}{\partial t}
\frac{\partial \eta}{\partial x}\right)\right\rangle
-\frac{1}{2}\frac{dU^{\prime\prime}(\phi_0)}{dx}\left\langle\eta^2
\right\rangle\,.
\end{align}
The first term on the right hand side vanishes by the field equation
for $\eta$. The second term is  the time derivative of $T_{01}$, which
by itself is zero. Hence Eq.~(\ref{eq:B3}) properly reflects the
continuity equation for $T_{\nu1}$, and thus corroborates the finite
part of the renormalization in Eq.~(\ref{eqn:quadratic2}).

\bibliographystyle{apsrev}
\bibliography{kinkdensity}

\end{document}